\documentclass[cameraready]{Interspeech}

\title{VoiceTTA: Enhancing Zero-Shot Text-to-Speech via Reinforcement Learning-Based Test-Time Adaptation}

\author[affiliation={1}]{Tianxin}{Xie}
\author[affiliation={2}]{Chenxing}{Li}
\author[affiliation={2}]{Dong}{Yu}
\author[affiliation={1},correspondingauthor]{Li}{Liu}

\address{
    $^1$ The Hong Kong University of Science and Technology (Guangzhou) \\
    $^2$ Tencent
}

\email{$^*$avrillliu@hkust-gz.edu.cn}


\usepackage{comment}
\usepackage{amsmath,graphicx,hyperref}
\usepackage{amssymb}
\usepackage{booktabs}
\usepackage{algorithm}
\usepackage{algorithmic}
\usepackage{xcolor}
\usepackage{threeparttable}
\usepackage{multirow}
\usepackage{makecell}
\usepackage[table]{xcolor}

\begin{document}

\maketitle

\begin{abstract}
Recently, zero-shot text-to-speech (TTS) has enabled high-fidelity and expressive speech synthesis, but it often fails to imitate unseen speaking styles from uncommon scenarios (e.g., crosstalk, dialects). Moreover, fine-tuning pretrained models requires large, high-quality datasets, limiting rapid personalization. We propose VoiceTTA, a reinforcement learning-based test-time adaptation (TTA) method that improves voice imitation of pretrained zero-shot TTS models. VoiceTTA introduces two style rewards based on coefficient-of-variation differences of F0 and energy, combined with speaker similarity and intelligibility (WER from a pretrained Whisper model), and optimizes learnable prefixes via group relative preference optimization (GRPO) in a flow matching-based model at inference time. Extensive experiments demonstrate substantial improvements on uncommon speech prompts, outperforming state-of-the-art baselines. Audio samples are available at \url{https://voicetta.pages.dev/}.
\end{abstract}

\section{Introduction}

Zero-shot text-to-speech (TTS)~\cite{kim2021vits,chen2025valle,xie2025towards} has advanced considerably in recent years, enabling users to generate authentic, natural-sounding personalized speech from a reference prompt.
However, existing zero-shot TTS models are primarily trained on large-scale datasets from common scenarios (e.g., podcasts, TV shows, and audiobooks)~\cite{he2024emilia,zen2019libritts,rong2025audiogenie}, which limits their effectiveness in less common settings, such as crosstalk, game commentary, and regional dialects.
This highlights the need for a lightweight and efficient adaptation method for such scenarios.

Existing zero-shot TTS methods encounter two key challenges when processing uncommon speech prompts.
\textbf{First}, such prompts (e.g., slurred, dialects) often feature accents, speaking styles, and linguistic content that differ substantially from the datasets on which these models are trained.
This mismatch introduces a domain shift between training and test data, causing generated speech to retain characteristics of the training domain while failing to capture the unique style of unseen prompts.
For example, VALL-E~\cite{chen2025valle} and its successors~\cite{zhang2023vallex,meng-etal-2025-autoregressive} can produce intelligible speech that matches the timbre of speech prompts, but often fail to imitate exaggerated prosody or regional accents.
\textbf{Second}, collecting and annotating speech datasets with dramatic vocal variations for uncommon scenarios is laborious, which compounds the difficulty of training high-quality TTS models for such cases.
Moreover, deployed TTS systems inevitably encounter corner cases that are not covered in training datasets.
Thus, these two issues impede the broader application of zero-shot TTS technologies.

\begin{figure}[t]
    \centering
    \includegraphics[width=\linewidth]{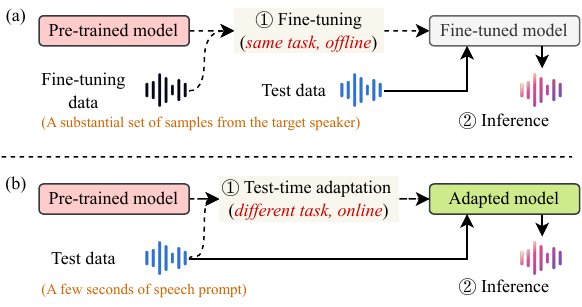}
    \caption{(a) The fine-tuning paradigm repeats the pre-training task on target-speaker data. (b) The TTA paradigm learns additional knowledge from the test data via a different task.}
    \label{fig:tta}
\end{figure}

Adapting pre-trained zero-shot TTS models to unseen speakers can help alleviate domain shift. Existing speaker adaptation methods mainly include: (1) speaker embedding-based adaptation, which learns speaker-specific representations (e.g., x-vectors~\cite{Snyder2018xvectors}) to enable synthesis for new speakers with little data, but relies on high-quality embedding models~\cite{casanova2022yourtts,lu2019one}; and (2) fine-tuning-based adaptation, which updates model parameters using more target-speaker data to better capture voice traits such as accent and style, though it is more time-consuming and data-intensive~\cite{Huang2024voicetuner,neekhara2021adapting}. However, both approaches may struggle to generalize to diverse speaking styles in unseen prompts and lack efficient adaptability to new users.


To address these challenges, we propose a reinforcement learning-based test-time adaptation (TTA~\cite{liang2025comprehensive}) framework to improve voice imitation in pre-trained zero-shot TTS models.
As shown in Fig.~\ref{fig:tta}, TTA performs lightweight parameter adaptation during inference using only a few speech samples from the target speaker, enabling efficient online adaptation to unseen prompts.
Compared with traditional fine-tuning that requires large-scale data, TTA can quickly align synthesized speech with the prosodic and stylistic characteristics of reference audio using only seconds of speech.

\begin{figure*}[t]
    \centering
    \includegraphics[width=\linewidth]{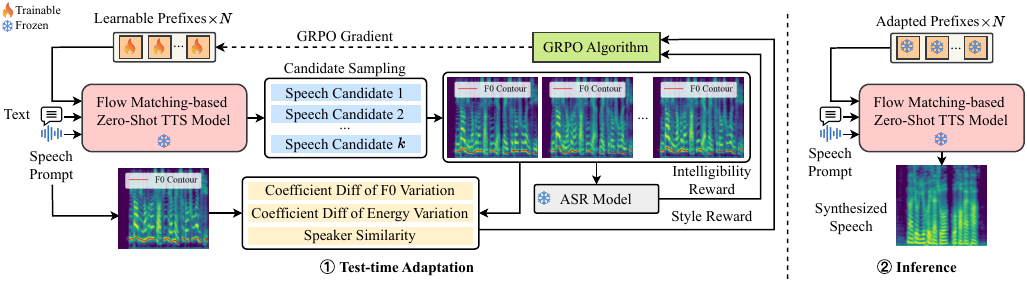}
    \caption{The overview of the proposed VoiceTTA method. We first apply the GRPO algorithm to optimize learnable prefixes in a flow matching-based TTS model, guided by intelligibility and style rewards. Then, we use the adapted prefixes to guide new speech synthesis.}
    \label{fig:framework}
\end{figure*}

A key challenge in TTA is the design of effective auxiliary tasks, since directly transferring tasks from other domains often leads to suboptimal performance~\cite{wang2021tent,sun2020test}.
Our experiments show that modeling F0 variation, energy variation, and speaker similarity as auxiliary tasks improves voice imitation, while incorporating an ASR-related objective helps maintain speech intelligibility.
During inference, multiple candidate utterances are generated, and auxiliary-task-based rewards are computed to train the group relative preference optimization (GRPO) algorithm~\cite{shao2024deepseekmath}, which optimizes lightweight learnable prefixes before final synthesis.
To our knowledge, this is the first work to enhance the voice imitation capability of pre-trained zero-shot TTS models on unseen prompts via TTA. In summary, our contributions are as follows:
\begin{itemize}
    \item We propose VoiceTTA, a reinforcement learning-based test-time adaptation framework that improves zero-shot TTS models by optimizing lightweight learnable prefixes with minimal parameter overhead.
    \item The proposed F0- and energy-variation rewards, together with the speaker similarity reward, guide the TTS model to capture acoustic details from the reference speech and enhance its imitation ability, while the WER-based reward improves voice intelligibility.
    \item Extensive objective and subjective evaluations on several SOTA zero-shot TTS models demonstrate the effectiveness of our method, particularly on uncommon prompts.
\end{itemize}

\section{Method}

\subsection{Overview}
Given an unseen speech prompt $x$ and text $p$, our goal is to maximize the probability that the speech $y=F_\theta(x,p)$ generated by a TTS model $F_\theta$ closely matches the speaking style of $x$ at inference time:
\begin{equation}
    \underset{\theta}{\textnormal{minimize:}}~~D(x,F_\theta(x,p)),
\end{equation}
where $D$ is a distance function that measures the similarity in speech characteristics between two audio samples.
However, $D$ cannot be explicitly defined and must be estimated in practice.
As shown in Fig.~\ref{fig:framework}, to minimize $D$ in the TTA setting, we employ GRPO with intelligibility and style rewards to optimize the parameters of the learnable prefixes, sampling $k$ candidate speeches for reward computation.
For each candidate, the intelligibility reward is measured by the Word Error Rate (WER), while the style reward is derived from the coefficient of variation of F0 (F0-CV), the coefficient of variation of energy (Energy-CV), and speaker similarity (S-SIM) between the speech prompt and the synthesized output.
After $G$ GRPO steps, the adapted model is used to synthesize new speech.
It is worth noting that the additional model parameters are minimal and cheap to store. Adaptation requires only a few seconds of user audio, after which the personalized model supports fast, repeated inference, making it well-suited for online deployment.

\subsection{Speech Candidate Sampling}

In this paper, we use a flow matching–based TTS model to sample speech candidates. The flow matching process transforms a latent variable into a mel-spectrogram conditioned on the encoded text and speech prompts, progressively transporting a prior distribution toward the learned data manifold in the acoustic feature space~\cite{lipman2023flow}. The latent variable is initially sampled from an isotropic Gaussian prior $\mathcal{N}(0, T^2 I)$, where the temperature $T$ controls the stochasticity of generation: higher $T$ increases prosodic and stylistic variation, while lower $T$ produces more stable but less diverse outputs.
The latent is then converted to a mel-spectrogram via the flow model and finally to a waveform using a vocoder~\cite{siuzdak2024vocos}.
To generate $k$ diverse candidates for reward computation, $T$ is sampled from a uniform distribution, while all candidates share the same text content.

\subsection{Rewards Computation}

We use $k$ candidates and a speech prompt to compute the style reward (F0-CV, Energy-CV, S-SIM) and the intelligibility reward (WER).
To measure the similarity in pitch dynamics between a candidate and the reference speech, we compute the F0-CV difference between them.
Let $F0_{\text{gen}} = [f_1^{\text{gen}}, \dots, f_{LG}^{\text{gen}}]$ and $F0_{\text{ref}} = [f_1^{\text{ref}}, \dots, f_{LR}^{\text{ref}}]$ denote the voiced frame-level F0 contours of the candidate and the reference speech, respectively.
We first compute their F0-CV as:
\begin{equation}
    \text{F0-CV}_\text{gen} = \frac{\sigma_{F0_{\text{gen}}}}{\mu_{F0_{\text{gen}}}}, \text{F0-CV}_\text{ref} = \frac{\sigma_{F0_{\text{ref}}}}{\mu_{F0_{\text{ref}}}},
\end{equation}
where $\mu$, $\sigma$ are the mean and standard deviation of the F0 contours, respectively.
The F0-CV reward is then defined as:
\begin{equation}
    r_\text{F0-CV} = -|\text{F0-CV}_\text{gen}-\text{F0-CV}_\text{ref}|.
\end{equation}
A higher $r_\text{F0-CV}$ indicates closer alignment with the reference pitch dynamics, positively contributing to the style reward.

To quantify the similarity in energy dynamics between a candidate and the reference speech, we extract voiced frame-level energy sequences $E_{\text{gen}} = [e_1^{\text{gen}}, \dots, e_{LG}^{\text{gen}}]$ and $E_{\text{ref}} = [e_1^{\text{ref}}, \dots, e_{LR}^{\text{ref}}]$, where the energy of each voiced frame is the sum of its mel frequencies:
\begin{equation}
    e_j = \sum_{m} M(j, m),
\end{equation}
where $M$ is the mel-spectrogram, $j$ denotes the frame index, $m$ denotes the Mel filter bank channel index.
The Energy-CV of the candidate and the reference speech are:
\begin{equation}
    \text{Energy-CV}_\text{gen} = \frac{\sigma_{E_{\text{gen}}}}{\mu_{E_{\text{gen}}}}, \text{Energy-CV}_\text{ref} = \frac{\sigma_{E_{\text{ref}}}}{\mu_{E_{\text{ref}}}},
\end{equation}
where $\mu$ and $\sigma$ denote the mean and standard deviation of the frame-level energy, respectively.
Then, the Energy-CV reward for a candidate is defined as:
\begin{equation}
    r_\text{Energy-CV} = -|\text{Energy-CV}_\text{gen}-\text{Energy-CV}_\text{ref}|.
\end{equation} 
A larger Energy-CV indicates that the generated speech better captures the energy dynamics of the reference speech.

The S-SIM reward between a candidate and the reference speech is computed as the cosine similarity between their embeddings extracted from a speaker embedding model $SE$:
\begin{equation}
    r_\text{S-SIM} = \cos(SE(A_\text{gen}), SE(A_\text{ref})),
\end{equation}
where $A_\text{gen}$ and $A_\text{ref}$ are the waveforms of the candidate and the reference speech.
The intelligibility reward $r_\text{Intel}$ for each candidate is computed as its WER.

\subsection{GRPO-Based Test-Time Adaptation}

To adapt the TTS model to unseen speech prompts, we treat the learnable prefixes as a stochastic policy $\pi_\theta$ within a reinforcement learning framework and employ GRPO, which needs no value model and computes rewards using rule-based or model-based methods, as implemented in our approach.
Assuming we have $k$ candidates and their rewards ($r^i_\text{F0-CV}, r^i_\text{Energy-CV}, r^i_\text{S-SIM}, r^i_\text{Intel}, i \in [1,k]$), we first normalize each type of reward into the range [0,1] for $k$ candidates.
The normalized rewards for each candidate are denoted as $s^i_\text{F0-CV}, s^i_\text{Energy-CV}, s^i_\text{S-SIM}, s^i_\text{Intel}$.
Then the total reward $r^i$ for candidate $i$ is the sum of the normalized rewards:
\begin{equation} \label{eq:final_reward}
    r^i = \lambda_1 s^i_\text{F0-CV} + \lambda_2 s^i_\text{Energy-CV} + \lambda_3 s^i_\text{S-SIM} + \lambda_4 s^i_\text{Intel}.
\end{equation}
Then, the GRPO loss for our method is defined as:
\begin{align}
    J_{GRPO} &= \mathbb{E} \left[ \sum_{i=1}^k \left( \min \left( \frac{\pi_\theta(o_i)}{\pi_{\theta_\text{old}}(o_i)} A_i,\right. \right. \right. \notag \\
    &\quad \left.\left.\left. \text{Clip}\left(\frac{\pi_\theta(o_i)}{\pi_{\theta_\text{old}}(o_i)}, 1-\epsilon, 1+\epsilon\right) A_i \right) \right) \right],
\end{align}
where
\begin{equation}
    A_i = \frac{r^i-\text{mean}(\{ r^1,r^2,...,r^k \})}{\text{std}(\{ r^1,r^2,...,r^k \})}.
\end{equation}
Note that there is no KL term in $J_{GRPO}$ because we are optimizing the lightweight learnable prefixes instead of the whole model, which is different from the original GRPO loss~\cite{shao2024deepseekmath}.
In our work, we prepend multiple learnable prefixes to the input of the first layer of a pre-trained TTS model.
Unlike LLMs, which predict token probabilities, flow-matching-based models directly regress mel-spectrograms.
Therefore, we adopt the flow matching loss as a probability proxy to compute $\pi_\theta(o_i)$ and $\pi_{\theta_\text{old}}(o_i)$, which measure the tendency to generate output $o_i$.
Following~\cite{lipman2023flow}, the flow matching loss is defined as:
\begin{equation}
    \mathcal{L}_{FM, \theta}(o) = \mathbb{E}_{t, p(x_t|o)} [\left|v_{\theta}(x_t, t) - u_t(x_t|o)\right|^2],
\end{equation}
which measures the discrepancy between the model-predicted velocity field $v_{\theta}(x_t, t)$ and the ideal target velocity field $u_t(x_t|o)$.
The lower the flow-matching loss for a sample $o$, the more accurately the model's learned dynamics can transform noise into $o$. Therefore, we can reasonably assume:
\begin{equation}
    \pi_\theta (o) = \log p_{\theta}(o) \propto -L_{\mathrm{FM},\theta}(o).
\end{equation}
Thus, we define the probability ratio term in GRPO as:
\begin{align}
\frac{\pi_\theta(o_i)}{\pi_{\theta_\text{old}}(o_i)} &= \log p_{\theta}(o_i) - \log p_{\theta_{\text{old}}}(o_i) \notag \\
          &\approx (-\mathcal{L}_{FM, \theta}(o_i)) - (-\mathcal{L}_{FM, \theta_{\text{old}}}(o_i)) \notag \\
          &= \mathcal{L}_{FM, \theta_{\text{old}}}(o_i) - \mathcal{L}_{FM, \theta}(o_i).
\end{align}
After $G$ GRPO adaptation steps on the unseen speech prompt, we use the model to synthesize new speech samples.
After adaptation and prediction for a given sample, the prefixes are randomly reinitialized before processing the next one, preventing updates from accumulating across the test set.

\section{Experiment}

\subsection{Experimental Settings}

\noindent\textbf{Datasets.} 
We collected an internal dataset of 200 speech samples with uncommon speaking styles (90 accented, 40 children’s, 30 slurred, and 40 from Chinese sketches), which are underrepresented in existing corpora and serve as realistic test-time scenarios. To further assess dialectal adaptation, we sampled 160 utterances (20 per dialect) from KeSpeech~\cite{tang2021kespeech}, covering eight Chinese dialects.

\noindent\textbf{Models.}
We use F5-TTS~\cite{chen-etal-2025-f5} as our backbone and the baseline. We also compare with three SOTA zero-shot TTS models, i.e., CosyVoice~\cite{du2024cosyvoice}, MaskGCT~\cite{wang2025maskgct}, and Vevo~\cite{zhang2025vevo}.

\noindent\textbf{Implementation.}
For each test sample, we generate four candidates to compute the GRPO reward.
For the final reward in Eq.~\ref{eq:final_reward}, we set $\lambda_1$=$\lambda_2$=0.2, $\lambda_3$=1, and $\lambda_4$=1.5, as $r_\text{S-SIM}$ and $r_\text{Intel}$ primarily drive speaker similarity and WER reduction, as demonstrated in the ablation study.
We prepend 4 learnable prefixes to the input of the first DiT layer of F5-TTS.
The learning rate is set to $5\times10^{-4}$ with a 5\% warmup ratio, and we run $G$=50 GRPO adaptation steps.
For candidate sampling, we draw a new $T$ from the uniform distribution $U(0.5, 1.5)$  to synthesize a new speech sample, and we sample 4 candidates for each GRPO optimization step.
All the evaluations are conducted on an NVIDIA RTX 6000 Ada GPU.

\noindent\textbf{Evaluation Metrics.}
For objective evaluation, we report WER and Speaker Similarity (S-SIM). WER is derived from Whisper-Large V3~\cite{radford2023robust} transcriptions, while S-SIM is computed as the cosine similarity between embeddings of the synthesized speech and a neutral reference using a speaker embedding model~\cite{Bredin2020}. For subjective evaluation, we report Naturalness MOS (N-MOS) and Similarity MOS (S-MOS), both rated on a 1–5 scale. N-MOS measures perceived naturalness (1 = very unnatural, 5 = completely natural), while S-MOS evaluates perceived style similarity (1 = very dissimilar, 5 = highly similar).

\begin{table}[!t]
\setlength{\tabcolsep}{3pt}
\renewcommand\arraystretch{0.9}
\caption{Performance comparison on five test-time scenarios.}
\centering
\scriptsize
\begin{threeparttable}
    \begin{tabular}{cccccc}
        \toprule
        & Method & WER($\downarrow$) & S-SIM($\uparrow$) & S-MOS($\uparrow$) & N-MOS($\uparrow$) \\
        \midrule
        \multirow{5}{*}{Accented} & CosyVoice~\cite{du2024cosyvoice} & 3.18 & 0.58 & 3.26 & \textbf{3.94} \\
        & MaskGCT~\cite{wang2025maskgct} & 3.09 & \textbf{0.69} & 3.18 & 3.52 \\ 
        & Vevo~\cite{zhang2025vevo} & 5.72 & 0.44 & 3.01 & 2.37 \\
        & F5-TTS~\cite{chen-etal-2025-f5} & \textbf{2.81} & 0.67 & 3.22 & 3.62 \\
        \cmidrule(rl){2-6}
        & VoiceTTA (Ours) & 2.82 & \textbf{0.69} & \textbf{3.46} & 3.51 \\
        \midrule
        \multirow{5}{*}{Children} & CosyVoice~\cite{du2024cosyvoice} & 3.26 & 0.53 & 3.34 & \textbf{3.85} \\
        & MaskGCT~\cite{wang2025maskgct} & 3.11 & 0.61 & 3.26 & 3.41 \\ 
        & Vevo~\cite{zhang2025vevo} & 9.54 & 0.45 & 3.05 & 2.11 \\
        & F5-TTS~\cite{chen-etal-2025-f5} & 3.03 & 0.60 & 3.19 &  3.51 \\
        \midrule
        & VoiceTTA (Ours) & \textbf{3.01} & \textbf{0.63} & \textbf{3.37} & 3.50 \\
        \midrule
        \multirow{5}{*}{Slurred} & CosyVoice~\cite{du2024cosyvoice} & 5.18 & 0.51 & 3.05 & \textbf{3.25} \\
        & MaskGCT~\cite{wang2025maskgct} & 4.82 & 0.57 & 3.09 & 2.98 \\
        & Vevo~\cite{zhang2025vevo} & 16.73 & 0.29 & 1.54 & 1.34 \\
        & F5-TTS~\cite{chen-etal-2025-f5} & 4.57 & 0.55 & 2.83 & 3.05 \\
        \cmidrule(rl){2-6}
        & VoiceTTA (Ours) & \textbf{4.49} & \textbf{0.58} & \textbf{3.44} & 3.07 \\
        \midrule
        \multirow{5}{*}{\makecell{Chinese\\ Sketches}} & CosyVoice~\cite{du2024cosyvoice} & 5.02 & 0.55 & 3.16 & \textbf{3.55} \\
        & MaskGCT~\cite{wang2025maskgct} & 4.73 & 0.49 & 3.08 & 3.17 \\
        & Vevo~\cite{zhang2025vevo} & 13.65 & 0.35 & 2.16 & 1.83 \\
        & F5-TTS~\cite{chen-etal-2025-f5} & \textbf{3.11} & 0.58 & 3.11 & 3.45 \\
        \cmidrule(rl){2-6}
        & VoiceTTA (Ours) & 3.26 & \textbf{0.60} & \textbf{3.25} & 3.44 \\
        \midrule
        \multirow{5}{*}{\makecell{Chinese\\ Dialects}} & CosyVoice~\cite{du2024cosyvoice} & 4.92 & 0.51 & 3.14 & \textbf{3.26} \\
        & MaskGCT~\cite{wang2025maskgct} & 4.16 & 0.56 & 3.10 & 3.01 \\
        & Vevo~\cite{zhang2025vevo} & 15.69 & 0.28 & 1.82 & 1.57 \\
        & F5-TTS~\cite{chen-etal-2025-f5} & 3.38 & 0.59 & 2.93 & 3.16 \\
        \cmidrule(rl){2-6}
        & VoiceTTA (Ours) & \textbf{3.13} & \textbf{0.62} & \textbf{3.18} & 3.20 \\
        \midrule
        \multirow{5}{*}{Averaged} & CosyVoice~\cite{du2024cosyvoice} & 4.57 & 0.54 & 3.25$\pm$0.92 & \textbf{3.58$\pm$0.73} \\
        & MaskGCT~\cite{wang2025maskgct} & 3.26 & 0.62 & 3.14$\pm$0.93 & 3.14$\pm$1.07 \\ 
        & Vevo~\cite{zhang2025vevo} & 12.41 & 0.34 & 2.05$\pm$1.03 & 1.91$\pm$1.01 \\
        & F5-TTS~\cite{chen-etal-2025-f5} & 3.19 & 0.57 & 3.07$\pm$1.07 & 3.36$\pm$1.04 \\
        \cmidrule(rl){2-6}
        \rowcolor{gray!20} & Ours & \textbf{3.12} & \textbf{0.64} & \textbf{3.27$\pm$0.62} & 3.35$\pm$0.77 \\
        \bottomrule
    \end{tabular}
\end{threeparttable}
\label{tab:eval}
\end{table}


\subsection{Objective Evaluation}

As shown in Table~\ref{tab:eval}, our method achieves a WER of 3.12, outperforming F5-TTS (3.19), MaskGCT (3.26), and substantially surpassing CosyVoice (4.57) and Vevo (12.41). These results demonstrate that VoiceTTA enhances stylistic attributes without sacrificing clarity, confirming that improvements in voice imitation do not come at the cost of intelligibility.
For S-SIM, our method attains the highest score (0.64), exceeding F5-TTS (0.57), MaskGCT (0.62), CosyVoice (0.54), and Vevo (0.34).
This improvement highlights the effectiveness of our reinforcement learning–based TTA framework in capturing target speaker characteristics, even under challenging prompts.

\subsection{Subjective Evaluation}

We evaluate perceptual style similarity using S-MOS.
24 participants are asked to rate S-MOS between 20 pairs of reference and generated samples.
Table~\ref{tab:eval} shows that our method achieves the highest score (3.27), slightly surpassing CosyVoice (3.25) and clearly outperforming F5-TTS (3.07), MaskGCT (3.14), and Vevo (2.05).
This aligns with the S-SIM results, where our method also ranks highest (0.64), confirming that style-based rewards effectively guide adaptation toward target characteristics.
For naturalness, our method attains an N-MOS of 3.35, comparable to F5-TTS (3.36) and slightly below CosyVoice (3.58), which is pretrained on over 130,000 hours and post-trained on 500 hours of high-quality speech.
While not the top score, this demonstrates that the improvements in style similarity do not compromise fluency or intelligibility.
Overall, VoiceTTA achieves superior stylistic fidelity while maintaining competitive naturalness, underscoring its balanced design.

\subsection{Ablation Study}

\textbf{Reward Types.}
We conduct an ablation study on the four reward types, evaluating S-SIM and WER using the same dataset and hyperparameters in the main experiment. As shown in Table~\ref{tab:ablation}, using only the intelligibility reward ($r_\text{Intel}$) achieves the lowest WER (3.10) but the weakest S-SIM (0.43), while combining the three style rewards ($r_\text{F0-CV}$ + $r_\text{Energy-CV}$ + $r_\text{S-SIM}$) yields the highest S-SIM (0.67) but severely degrades intelligibility (WER = 7.04). The full model, integrating all four rewards, balances both objectives, reaching a WER of 3.12 and an S-SIM of 0.64. These results highlight the complementary roles of these rewards and the necessity of a composite reward for effective style imitation without sacrificing intelligibility.

\begin{figure}[t]
    \centering
    \includegraphics[width=0.9\linewidth]{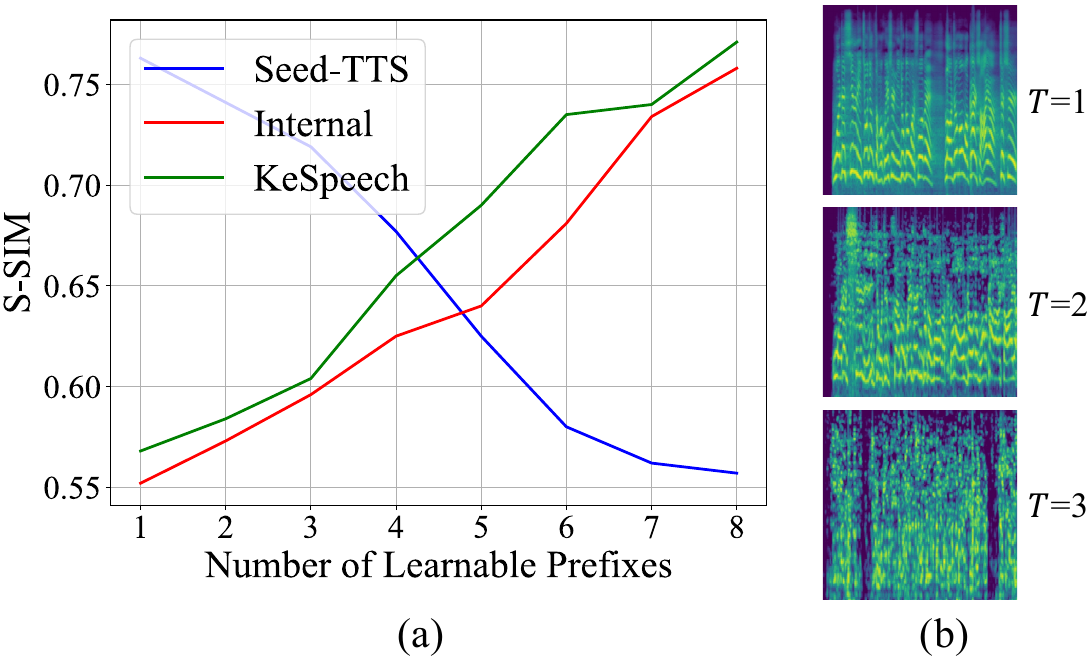}
    \caption{(a) The effect of the number of learnable prefixes. (b) High temperature $T$ reduces intelligibility.}
    \label{fig:prefix_temperature}
\end{figure}

\noindent\textbf{Number of Learnable Prefixes.}
We analyze the effect of the number of learnable prefixes by reporting average S-SIM on the Seed-TTS~\cite{anastassiou2024seed} test-en set using models adapted on KeSpeech and our internal dataset.
As shown in Fig.~\ref{fig:prefix_temperature}-a, S-SIM on Seed-TTS peaks with a single prefix and declines as more are added, suggesting that excessive adaptation may harm synthesis.
In contrast, on KeSpeech and our internal dataset, S-SIM improves with more prefixes, indicating enhanced adaptability to novel characteristics.
We therefore use four prefixes in the main experiments to balance adaptation and in-domain performance.

\noindent\textbf{Temperature $T$.}
As shown in Fig.~\ref{fig:prefix_temperature}-b, large $T$ values yield unintelligible speech. Therefore, we sample $T$ from a uniform distribution with a small range as in the experimental setting.

\begin{table}[!t]
    \caption{Ablation study on the four reward types.}
    \centering
    \scriptsize
    \renewcommand\arraystretch{0.9}
    \begin{tabular}{ccc}
        \toprule
        Reward & WER($\downarrow$) & S-SIM($\uparrow$) \\
        \midrule
        $r_\text{F0-CV}$ & 7.62 & 0.53 \\
        $r_\text{Energy-CV}$ & 7.76 & 0.51 \\
        $r_\text{S-SIM}$ & 6.58 & 0.62 \\
        $r_\text{Intel}$ & \textbf{3.10} & 0.43 \\
        \midrule
        $r_\text{F0-CV}$ + $r_\text{Energy-CV}$ & 7.69 & 0.55 \\
        \midrule
        $r_\text{F0-CV}$ + $r_\text{Energy-CV}$ + $r_\text{S-SIM}$ & 7.04 & \textbf{0.67} \\
        \midrule
        \rowcolor{gray!20} $r_\text{F0-CV}$ + $r_\text{Energy-CV}$ + $r_\text{S-SIM}$ + $r_\text{Intel}$ & \textbf{3.12} & \textbf{0.64} \\
        \bottomrule
    \end{tabular}
    \label{tab:ablation}
\end{table}

\subsection{Model Parameters}

For each unseen speaker, only a lightweight 16 KB prompt needs to be stored, enabling efficient personalization.
This design makes VoiceTTA more practical and scalable than existing methods for real-world deployment, where rapid adaptation to new speakers is essential.

\section{Conclusion}

We present VoiceTTA, a reinforcement learning–based TTA method for flow matching–based TTS models that optimizes lightweight prefixes with multiple rewards to balance style similarity and intelligibility.
VoiceTTA surpasses SOTA zero-shot TTS models in style similarity on five uncommon test-time scenarios while preserving naturalness, requires only a one-time adaptation, and introduces negligible parameters, making it practical for online, real-world personalized speech synthesis.

\section{Acknowledgement}

This work was supported by the National Natural Science Foundation of China (No. 62471420), GuangDong Basic and Applied Basic Research Foundation (2025A1515012296), and 2025 Tencent AI Lab Rhino-Bird Program.

\section{Generative AI Use Disclosure}

We used ChatGPT only for language editing and stylistic refinement; no technical content was generated by the tool.

\bibliographystyle{IEEEtran}
\bibliography{mybib}

@inproceedings{chen-etal-2025-f5,
  title={{F5-TTS}: A fairytaler that fakes fluent and faithful speech with flow matching},
  author={Chen, Yushen and Niu, Zhikang and Ma, Ziyang and Deng, Keqi and Wang, Chunhui and JianZhao, JianZhao and Yu, Kai and Chen, Xie},
  booktitle={Proceedings of the 63rd Annual Meeting of the Association for Computational Linguistics (Volume 1: Long Papers)},
  pages={6255--6271},
  year={2025}
}

@inproceedings{
wang2025maskgct,
title={Mask{GCT}: Zero-Shot Text-to-Speech with Masked Generative Codec Transformer},
author={Yuancheng Wang and Haoyue Zhan and Liwei Liu and Ruihong Zeng and Haotian Guo and Jiachen Zheng and Qiang Zhang and Xueyao Zhang and Shunsi Zhang and Zhizheng Wu},
booktitle={The Thirteenth International Conference on Learning Representations},
year={2025},
  pages={1--24},
}

@article{zhang2023vallex,
  title={Speak foreign languages with your own voice: Cross-lingual neural codec language modeling},
  author={Zhang, Ziqiang and Zhou, Long and Wang, Chengyi and Chen, Sanyuan and Wu, Yu and Liu, Shujie and Chen, Zhuo and Liu, Yanqing and Wang, Huaming and Li, Jinyu and others},
  journal={arXiv preprint arXiv:2303.03926},
  year={2023}
}

@inproceedings{he2024emilia,
  title={Emilia: An extensive, multilingual, and diverse speech dataset for large-scale speech generation},
  author={He, Haorui and Shang, Zengqiang and Wang, Chaoren and Li, Xuyuan and Gu, Yicheng and Hua, Hua and Liu, Liwei and Yang, Chen and Li, Jiaqi and Shi, Peiyang and others},
  booktitle={2024 IEEE Spoken Language Technology Workshop (SLT)},
  pages={885--890},
  year={2024},
}

@INPROCEEDINGS{Snyder2018xvectors,
  author={Snyder, David and Garcia-Romero, Daniel and Sell, Gregory and Povey, Daniel and Khudanpur, Sanjeev},
  booktitle={ICASSP}, 
  title={X-Vectors: Robust DNN Embeddings for Speaker Recognition}, 
  year={2018},
  volume={},
  number={},
  pages={5329-5333},
}

@article{neekhara2021adapting,
  title={Adapting {TTS} models for new speakers using transfer learning},
  author={Neekhara, Paarth and Li, Jason and Ginsburg, Boris},
  journal={arXiv preprint arXiv:2110.05798},
  year={2021}
}

@inproceedings{lu2019one,
  title={One-shot voice conversion with global speaker embeddings},
  author={Lu, Hui and Wu, Zhiyong and Dai, Dongyang and Li, Runnan and Kang, Shiyin and Jia, Jia and Meng, Helen},
  booktitle={Interspeech},
  pages={669--673},
  year={2019}
}

@inproceedings{Huang2024voicetuner,
  title={{VoiceTuner}: Self-Supervised Pre-training and Efficient Fine-tuning For Voice Generation},
  author={Huang, Rongjie and Wang, Yongqi and Hu, Ruofan and Xu, Xiaoshan and Hong, Zhiqing and Yang, Dongchao and Cheng, Xize and Wang, Zehan and Jiang, Ziyue and Ye, Zhenhui and others},
  booktitle={Proceedings of the 32nd ACM International Conference on Multimedia},
  pages={10630--10639},
  year={2024}
}

@inproceedings{meng-etal-2025-autoregressive,
  title={Autoregressive speech synthesis without vector quantization},
  author={Meng, Lingwei and Zhou, Long and Liu, Shujie and Chen, Sanyuan and Han, Bing and Hu, Shujie and Liu, Yanqing and Li, Jinyu and Zhao, Sheng and Wu, Xixin and others},
  booktitle={Proceedings of the 63rd Annual Meeting of the Association for Computational Linguistics (Volume 1: Long Papers)},
  pages={1287--1300},
  year={2025}
}

@inproceedings{xie2025towards,
  title={Towards controllable speech synthesis in the era of large language models: A systematic survey},
  author={Xie, Tianxin and Rong, Yan and Zhang, Pengfei and Wang, Wenwu and Liu, Li},
  booktitle={Proceedings of the 2025 Conference on Empirical Methods in Natural Language Processing},
  pages={764--791},
  year={2025}
}

@inproceedings{zen2019libritts,
  author       = {Heiga Zen and
                  Viet Dang and
                  Rob Clark and
                  Yu Zhang and
                  Ron J. Weiss and
                  Ye Jia and
                  Zhifeng Chen and
                  Yonghui Wu},
  title        = {Libri{TTS}: {A} Corpus Derived from LibriSpeech for Text-to-Speech},
  booktitle    = {Interspeech},
  pages        = {1526--1530},
  year         = {2019},
}

@inproceedings{tang2021kespeech,
  title={Kespeech: An open source speech dataset of mandarin and its eight subdialects},
  author={Tang, Zhiyuan and Wang, Dong and Xu, Yanguang and Sun, Jianwei and Lei, Xiaoning and Zhao, Shuaijiang and Wen, Cheng and Tan, Xingjun and Xie, Chuandong and Zhou, Shuran and others},
  booktitle={Thirty-fifth conference on neural information processing systems datasets and benchmarks track (Round 2)},
  year={2021}
}

@inproceedings{sun2020test,
  title={Test-time training with self-supervision for generalization under distribution shifts},
  author={Sun, Yu and Wang, Xiaolong and Liu, Zhuang and Miller, John and Efros, Alexei and Hardt, Moritz},
  booktitle={International conference on machine learning},
  pages={9229--9248},
  year={2020},
}

@inproceedings{
wang2021tent,
title={Tent: Fully Test-Time Adaptation by Entropy Minimization},
author={Dequan Wang and Evan Shelhamer and Shaoteng Liu and Bruno Olshausen and Trevor Darrell},
booktitle={International Conference on Learning Representations},
year={2021},
  pages={1--15},
}

@inproceedings{kim2021vits,
  title={Conditional variational autoencoder with adversarial learning for end-to-end text-to-speech},
  author={Kim, Jaehyeon and Kong, Jungil and Son, Juhee},
  booktitle={International conference on machine learning},
  pages={5530--5540},
  year={2021},
}

@inproceedings{
zhang2025vevo,
title={Vevo: Controllable Zero-Shot Voice Imitation with Self-Supervised Disentanglement},
author={Xueyao Zhang and Xiaohui Zhang and Kainan Peng and Zhenyu Tang and Vimal Manohar and Yingru Liu and Jeff Hwang and Dangna Li and Yuhao Wang and Julian Chan and Yuan Huang and Zhizheng Wu and Mingbo Ma},
booktitle={The Thirteenth International Conference on Learning Representations},
year={2025},
pages={1--24}
}

@ARTICLE{chen2025valle,
  author={Chen, Sanyuan and Wang, Chengyi and Wu, Yu and Zhang, Ziqiang and Zhou, Long and Liu, Shujie and Chen, Zhuo and Liu, Yanqing and Wang, Huaming and Li, Jinyu and He, Lei and Zhao, Sheng and Wei, Furu},
  journal={IEEE Transactions on Audio, Speech and Language Processing}, 
  title={Neural Codec Language Models are Zero-Shot Text to Speech Synthesizers}, 
  year={2025},
  volume={33},
  number={},
  pages={705-718},
}

@article{du2024cosyvoice,
  title={Cosyvoice: A scalable multilingual zero-shot text-to-speech synthesizer based on supervised semantic tokens},
  author={Du, Zhihao and Chen, Qian and Zhang, Shiliang and Hu, Kai and Lu, Heng and Yang, Yexin and Hu, Hangrui and Zheng, Siqi and Gu, Yue and Ma, Ziyang and others},
  journal={arXiv preprint arXiv:2407.05407},
  year={2024}
}

@inproceedings{radford2023robust,
  title={Robust speech recognition via large-scale weak supervision},
  author={Radford, Alec and Kim, Jong Wook and Xu, Tao and Brockman, Greg and McLeavey, Christine and Sutskever, Ilya},
  booktitle={International conference on machine learning},
  pages={28492--28518},
  year={2023},
}

@article{anastassiou2024seed,
  title={Seed-tts: A family of high-quality versatile speech generation models},
  author={Anastassiou, Philip and Chen, Jiawei and Chen, Jitong and Chen, Yuanzhe and Chen, Zhuo and Chen, Ziyi and Cong, Jian and Deng, Lelai and Ding, Chuang and Gao, Lu and others},
  journal={arXiv preprint arXiv:2406.02430},
  year={2024}
}

@inproceedings{Bredin2020,
  title={Pyannote. audio: neural building blocks for speaker diarization},
  author={Bredin, Herv{\'e} and Yin, Ruiqing and Coria, Juan Manuel and Gelly, Gregory and Korshunov, Pavel and Lavechin, Marvin and Fustes, Diego and Titeux, Hadrien and Bouaziz, Wassim and Gill, Marie-Philippe},
  booktitle={ICASSP 2020-2020 IEEE International conference on acoustics, speech and signal processing (ICASSP)},
  pages={7124--7128},
  year={2020},
}

@article{liang2025comprehensive,
  title={A comprehensive survey on test-time adaptation under distribution shifts},
  author={Liang, Jian and He, Ran and Tan, Tieniu},
  journal={International Journal of Computer Vision},
  volume={133},
  number={1},
  pages={31--64},
  year={2025},
}

@inproceedings{
siuzdak2024vocos,
title={Vocos: Closing the gap between time-domain and Fourier-based neural vocoders for high-quality audio synthesis},
author={Hubert Siuzdak},
booktitle={The Twelfth International Conference on Learning Representations},
year={2024},
  pages={1--15},
}

@inproceedings{rong2025audiogenie,
  title={Audiogenie: A training-free multi-agent framework for diverse multimodality-to-multiaudio generation},
  author={Rong, Yan and Wang, Jinting and Lei, Guangzhi and Yang, Shan and Liu, Li},
  booktitle={Proceedings of the 33rd ACM International Conference on Multimedia},
  pages={8872--8881},
  year={2025}
}

@inproceedings{
lipman2023flow,
title={Flow Matching for Generative Modeling},
author={Yaron Lipman and Ricky T. Q. Chen and Heli Ben-Hamu and Maximilian Nickel and Matthew Le},
booktitle={The Eleventh International Conference on Learning Representations },
year={2023},
pages={1--28}
}

@inproceedings{casanova2022yourtts,
  title={Yourtts: Towards zero-shot multi-speaker tts and zero-shot voice conversion for everyone},
  author={Casanova, Edresson and Weber, Julian and Shulby, Christopher D and Junior, Arnaldo Candido and G{\"o}lge, Eren and Ponti, Moacir A},
  booktitle={International conference on machine learning},
  pages={2709--2720},
  year={2022},
}

@article{shao2024deepseekmath,
  title={Deepseekmath: Pushing the limits of mathematical reasoning in open language models},
  author={Shao, Zhihong and Wang, Peiyi and Zhu, Qihao and Xu, Runxin and Song, Junxiao and Bi, Xiao and Zhang, Haowei and Zhang, Mingchuan and Li, YK and Wu, Yang and others},
  journal={arXiv preprint arXiv:2402.03300},
  year={2024}
}

\end{document}